\documentclass[11pt,aps,nofootinbib,superscriptaddress,amssymb]{revtex4}

\usepackage{amsmath,setspace,subfigure,amsfonts,latexsym}
\usepackage{amssymb}
\usepackage{color}
\usepackage{epsfig}
\usepackage{color}
\definecolor{White}{rgb}{1,1,1}
\definecolor{Red}{rgb}{1,0.1,0}
\definecolor{LightYellow}{rgb}{1,1,.875}
\definecolor{SteelBlue}{rgb}{.273,.508,.703}
\definecolor{navy}{rgb}{0,0,.5}
\definecolor{LightCyan}{rgb}{.875,1,1}
\definecolor{DarkRed}{rgb}{.543,0,0}
\definecolor{HotPink}{rgb}{1,.41,.70}
\definecolor{ForestGreen}{rgb}{.13,.54,.13}
\definecolor{OliveDrab}{rgb}{.42,.55,.14}
\definecolor{MediumBlue}{rgb}{0,0,.80}
\definecolor{RoyalBlue}{rgb}{.25,.41,.88}
\definecolor{DeepSkyBlue}{rgb}{0,.746,1}
\definecolor{Brown}{rgb}{0.545,0.271,0.074}

\def\bea{\begin{eqnarray}}
\def\eea{\end{eqnarray}}
\def\bec{\begin{center}}
\def\ec{\end{center}}

\def\beq{\begin{equation}}
\def\eeq{\end{equation}}

\newcommand\lsim{\mathrel{\rlap{\lower4pt\hbox{\hskip1pt$\sim$}}
    \raise1pt\hbox{$<$}}}
\newcommand\gsim{\mathrel{\rlap{\lower4pt\hbox{\hskip1pt$\sim$}}
    \raise1pt\hbox{$>$}}}
\def\bea{\begin{eqnarray}}
\def\eea{\end{eqnarray}}
\def\ba{\begin{array}}
\def\ea{\end{array}}
\def\bc{\begin{center}}
\def\ec{\end{center}}

\begin{document}

\hfill CTPU-16-44 

\hfill UT-16-37

\hfill IPMU-16-0190

\title{\Large  Affleck-Dine Leptogenesis with Varying Peccei-Quinn Scale}
\author{Kyu Jung Bae} 
\affiliation{Center for Theoretical Physics of the Universe, Institute for Basic Science (IBS), Daejeon 34501, Korea}
\author{Howard Baer}
\affiliation{Department of Physics and Astronomy, University of Oklahoma, Norman, OK 73019, USA}
\author{Koichi Hamaguchi}
\author{Kazunori Nakayama}
\affiliation{Department of Physics, University of Tokyo, Bunkyo-ku, Tokyo 113-0033, Japan}
\affiliation{Kavli Institute for the Physics and Mathematics of the Universe (Kavli IPMU), University of Tokyo, Kashiwa 277-8583, Japan}

\date{\today}

\begin{abstract}
The Affleck-Dine leptogenesis scenario along the $LH_u$ flat direction 
is reconsidered.
It is known that successful Affleck-Dine leptogenesis requires that the lightest neutrino mass is extremely small.
This situation can be significantly relaxed if the neutrino mass in the early universe is different from the present one.
We consider a supersymmetric Dine-Fischler-Srednicki-Zhitnitsky (DFSZ) type model, 
which provides a solution to the strong $CP$ problem and generates 
a SUSY $\mu$-term and right-handed neutrino masses.
If the PQ scale during lepton number generation is much larger than the present value, leptogenesis is very efficient so that enough baryon number can be generated without introducing a hierarchically small neutrino mass.
The final baryon asymmetry is related to the $\mu$-term, and hence linked to the level of electroweak fine--tuning.
We also show the PQ breaking scalar dynamics that keeps a large PQ breaking scale during inflation and lepton number generation. The $\mu$-term generating superpotential plays an important role for preserving the lepton asymmetry during saxion oscillation.
In this scenario, the axion isocurvature perturbation is naturally suppressed.

\end{abstract}

\maketitle

\titlepage

\section{Introduction}

The baryon asymmetry of the universe (BAU) is one of the intriguing puzzles in our understanding of the universe.
In order to generate the BAU, three conditions, which have been pointed out by Sakharov~\cite{Sakharov:1967dj}, 
need to be satisfied in the early universe: baryon number ($B$) violation, $C$/$CP$ violation and departure from equilibrium.
The $B$-violation is actually replaced by $(B-L)$-violation since the 
$(B-L)$-conserving and $(B+L)$-violating sphaleron process can easily transfer $B$ number into $L$ number and {\it vice versa} before the electroweak phase transition~\cite{Kuzmin:1985mm}.
While several ideas have been proposed to produce BAU by using $B$-violating interactions, 
leptogenesis provides a different mechanism for generating BAU by using $L$-violating interactions~\cite{Fukugita:1986hr}.
In such a case, instead of a baryon asymmetry, 
a lepton asymmetry is generated in the early universe via $L$-violating 
interactions.
Once the net lepton number is generated, it is transferred into the baryon number via the sphaleron process.

Such scenarios are naturally realized in seesaw models for neutrino mass generation.
The $L$-violating and $C$/$CP$-violating interactions are introduced for
Majorana mass terms and their Yukawa couplings of right-handed neutrinos (RHNs).
One of the most beautiful scenarios is the thermal leptogenesis that 
produces lepton asymmetry from thermally produced RHNs.
The thermal plasma after reheating produces a number of RHNs.
When the produced RHNs are out of equilibrium, they decay asymmetrically into leptons vs. anti-leptons and thus generate net lepton number.
It has been known that in this case, a large reheating temperature $T_R\gtrsim 10^9$ GeV 
is required to obtain enough baryon-number-to-entropy ratio $n_B/s\sim 10^{-10}$~\cite{Buchmuller:2005eh}.


If supersymmetry (SUSY) exists, however, one must take care of gravitino production for such a high reheating temperature~\cite{Kawasaki:2008qe}.
The Affleck-Dine (AD) mechanism~\cite{Affleck:1984fy,Dine:1995kz} is an attractive baryogenesis scenario in SUSY models
that does not necessarily require high reheating temperature and hence the tension with gravitino production is circumvented.\footnote{
	There are also models with nonthermal leptogenesis~\cite{Lazarides:1991wu,Asaka:1999yd,Hamaguchi:2001gw}
	that works for $T_R\gtrsim 10^{6}$\,GeV.
}
In SUSY models, it is quite generic to have many flat directions, so the baryon asymmetry can be efficiently generated along such flat directions. 
In particular, if one considers the $LH_u$ flat direction, whose non-renormalizable potential is generated by integrating out RHNs, 
sufficient lepton asymmetry can be produced at relatively small reheat temperature.
In this case, however, one difficulty arises. For the desired baryon asymmetry,the $LH_u$ direction must be very flat during lepton number generation, which results in a hierarchically light neutrino~\cite{Asaka:2000nb,Fujii:2001zr}.
While this is not excluded since currently there is no lower bound on the lightest neutrino mass, such a hierarchical structure of neutrino masses 
may not be natural from the viewpoint of model building.



One reason for a hierarchically small neutrino mass in the AD mechanism
stems from the fact that the RHN mass scale is responsible for both leptogenesis and neutrino mass generation.
Thus the flatness of the $LH_u$ direction is directly related to the lightness of neutrino.
One way to relax this constraint is to make the RHN mass dynamical.
If the RHN mass scale is very large during lepton number generation but become small during the current universe, 
the lightest neutrino can have a rather large mass.

As pointed out in~\cite{Asaka:2000nb}, 
such a dynamical RHN mass can be realized in a Peccei-Quinn (PQ) symmetric model
to solve the strong CP problem~\cite{Peccei:1977hh}.
If the RHNs are charged under PQ symmetry, their masses are generated by the PQ symmetry breaking~\cite{Langacker:1986rj}.
Thus, the RHN mass is time-dependent and it is determined by the dynamics of 
the PQ field.
During inflation, the Hubble-induced SUSY breaking potential holds the 
PQ field at the Planck scale, and thus PQ symmetry is broken at the Planck scale.
After inflation and lepton number generation, the Hubble-induced SUSY breaking effect becomes as small as ordinary SUSY breaking contributions, so the PQ field starts to oscillate and settles down at the current value of the PQ breaking scale.
The RHNs also have Planck scale masses from the inflation era to the 
lepton number generation era, and become lighter afterwards.
Therefore, it is possible to simultaneously accommodate very efficient AD leptogenesis when the PQ scale is around Planck scale and rather larger 
neutrino mass in the current universe.

In this work, we revisit the scenario of AD leptogenesis with a 
varying PQ scale, and provide its concrete realization in the Dine-Fischler-Srednicki-Zhitnitsky (DFSZ) type model~\cite{Dine:1981rt} .
In this model, the PQ breaking scalar fields provide a solution to the strong CP problem, generate RHN masses and the superpotential 
$\mu$-term for the Higgsino sector. 
In addition, the superpotential term responsible for the $\mu$-term plays an important role to maintain the generated lepton asymmetry during PQ field oscillation.
The isocurvature contribution from axion dark matter is naturally suppressed by the large PQ scale during inflation.


In this case, one additional crucial feature resides in the PQ field decay. 
As stated above, once the Hubble-induced SUSY breaking effect becomes comparable to the ordinary SUSY breaking terms, the PQ breaking field, 
which is called saxion, 
starts to oscillate with respect to the current PQ scale.
The saxion dominates the universe right after the reheating process, so its decay produces large amount of entropy.
Hence the final baryon asymmetry is sensitive to the saxion decay,
which depends on how the PQ sector couples to the standard model particles.
In the DFSZ case, 
the saxion decay is dependent on the $\mu$-term.
 Therefore the resulting baryon asymmetry after saxion decay is linked to 
electroweak fine-tuning, which is determined by the $\mu$-term.

In Sec.~\ref{sec:AD}, we analyze AD leptogenesis along the $LH_u$ flat 
direction with a dynamical PQ breaking scale and then estimate the baryon asymmetry
taking account of the dilution from saxion decay.
In Sec.~\ref{sec:PQ_dyn}, we investigate the dynamics of PQ breaking scalar fields and then examine conservation of lepton asymmetry during the saxion oscillation.
In Sec.~\ref{sec:susy_scale} we present the main results in the form of contours
of required $m_{\nu1}$ values in the $\mu$ vs. $f_a$ plane, which show a relation between the baryon asymmetry and the electroweak fine-tuning.
In Sec.~\ref{sec:PQ}, we discuss some cosmological implications 
of the PQ sector: the axion isocurvature perturbation and axino production.
We conclude in Sec.~\ref{sec:conc}.

\section{Affleck-Dine leptogenesis with dynamical Peccei-Quinn symmetry breaking} \label{sec:AD}

In this section, we show how dynamical PQ symmetry breaking accommodates enough baryon asymmetry and a relatively large neutrino mass.

\subsection{The model}

We consider the neutrino sector for AD leptogenesis and the 
PQ breaking sector, which are described by the following superpotential,
\begin{align}
&W=W_{\rm AD} + W_{\rm PQ}, \\
&W_{\rm AD} =\frac12 \lambda XNN + y_{\nu} NLH_u, ~~~~W_{\rm PQ} = \eta Z(XY-f^2) + \frac{g_\mu Y^2}{M_P}H_u H_d,
\label{eq:lagPQ}
\end{align}
where $N$ is the RHN, $L$ is the lepton doublet, $H_u$ ($H_d$) is the up-type  (down-type) Higgs doublet, $X$ and $Y$ are the PQ fields, $Z$ is a singlet scalar,
$\lambda$, $\eta$ and $g_\mu$ represent numerical coefficients, $f$ denotes the (present) PQ breaking scale and $M_P$ the reduced Planck scale.
The PQ charges and lepton numbers of these fields are given in 
Table~\ref{table:PQcharge}.

\begin{table}
\begin{center}
\begin{tabular}{c|c|c|c|c|c|c|c|}
& $X$ & $Y$ & $Z$ & $N$ & $L$ & $H_u$ & $H_d$ \\
\hline
\hline
{\rm PQ} & $-2$ & 2 & 0 & 1 & 1 & $-2$ & $-2$ \\ \hline
{\rm L} & 0 & 0 & 0 & $-1$ & 1 & $0$ & 0 
\end{tabular}
\caption{U(1) charges of the fields.}
\label{table:PQcharge}
\end{center}
\end{table}

When $X$ obtains a large field value, the RHN becomes massive and can be integrated out to obtain the effective superpotential:
\begin{equation}
W_{\rm AD, eff}=-\frac12\frac{y_{\nu}^2(LH_u)^2}{\lambda X}.
\label{eq:ADeff}
\end{equation}
The neutrino mass is generated by the see-saw mechanism at low energy $\left<X\right>\sim f$:
\begin{equation}
m_{\nu 1}=\frac{y_{\nu}^2\langle H_u\rangle^2}{\lambda f}\simeq \frac{v^2}{M_{\rm eff}},
\end{equation}
where $v\simeq 174$ GeV and 
\begin{equation}
M_{\rm eff}=\frac{v^2}{m_{\nu1}}=3.0\times10^{17}\mbox{ GeV}\left(\frac{10^{-4}\mbox{ eV}}{m_{\nu1}}\right).
\end{equation}
Here and in the following, we assume that the lepton asymmetry is generated along the flattest $LH_u$ direction, which corresponds to the smallest neutrino mass $m_{\nu_1}$.
As will become clear, $X$ is different from its low-energy vacuum expectation value (VEV) $f$ in the early universe.\footnote{For simplicity, we assume that $X=Y=f$ at the present universe.}
In particular, $X$ is shown to be fixed at $X=X_0 (\sim M_P)$
until the saxion begins to oscillate (see Sec.~\ref{sec:PQ_dyn}). 
Thus one can also define an effective scale in the early universe as 
\begin{equation}
\frac{y_{\nu}^2}{\lambda X_0}=\frac{1}{M_{\rm eff}}\left(\frac{f}{X_0}\right)\equiv\frac{1}{M_*}.
\label{eq:M_*}
\end{equation}
In a different way, $M_*$ is expressed by
\begin{equation}
M_*
=7.2\times10^{23}\mbox{GeV}\left(\frac{10^{-4}\mbox{ eV}}{m_{\nu1}}\right)
\left(\frac{10^{12}\mbox{ GeV}}{f}\right)
\left(\frac{X_0}{M_P}\right).
\end{equation}
Thus $M_*$ is much larger than $M_{\rm eff}$, which modifies the ordinary relation between low-energy neutrino mass
and the efficiency of AD leptogenesis~\cite{Asaka:2000nb}.

In what follows, we consider the dynamics of the AD field $\phi$, which parameterizes the $LH_u$ $D$-flat direction as
\begin{equation}
	L = \frac{1}{\sqrt 2}\begin{pmatrix} \phi \\ 0 \end{pmatrix}, \qquad
	H_u  = \frac{1}{\sqrt 2}\begin{pmatrix} 0\\ \phi \end{pmatrix}.
\end{equation}
The scalar potential of the AD field in the model of Eq.~\ref{eq:lagPQ} reads
\begin{align}
	&V_{\rm AD} = V_{F} + V_{\rm soft} + V_H,\\ 
	&V_F = \frac{y_\nu^4}{4\lambda^2}\frac{|\phi|^6}{|X|^2},
	\qquad V_{\rm soft} = m_\phi^2|\phi|^2 + \left[ a_m m_{\rm soft} \frac{y_{\nu}^2}{8\lambda}\frac{\phi^4}{X}+{\rm h.c.}\right], \\
	& V_H = -c_H H^2|\phi|^2 + \left[ a_H H \frac{y_{\nu}^2}{8\lambda}\frac{\phi^4}{X}+{\rm h.c.}\right],
\end{align}
where $V_{\rm soft}$ denotes the contribution from the soft SUSY breaking and $V_H$ the Hubble-induced terms.
Here $m_\phi$ and $m_{\rm soft}$ are of the soft mass scale 
($m_{\rm soft}\sim$ few TeV), 
$H$ is the Hubble parameter, $c_H (>0)$, $a_m$ and $a_H$ are $\mathcal O(1)$ coefficients.\footnote{
	See Ref.~\cite{Kasuya:2008xp} for the issue related to the existence of the Hubble-induced $A$ term.
	In this paper we simply assume that there is a Hubble-induced $A$ term.
}
For the moment, we assume that the PQ field $X$ is fixed at $X_0$ until the AD field begins to oscillate.
We examine the dynamics including the PQ field in Sec.~\ref{sec:PQ_dyn}.

From the Hubble induced mass and $V_F$, the AD field is stabilized at
\begin{equation}
|\phi|=\phi_0\simeq \sqrt{M_*H},
\end{equation}
for $H \gg m_\phi$. Writing
\begin{equation}
X=X_0e^{ia_X/X_0},\qquad \phi=\phi_0e^{ia_{\phi}/\phi_0},\qquad
\end{equation}
one finds
\begin{equation}
V_H \supset \frac14|a_H| H^3M_*\cos\left(-\frac{a_X}{X_0}+\frac{4a_{\phi}}{\phi_0}+\delta_{a_H}\right)
\end{equation}
where $\delta_{a_H}=\arg(a_H)$.
The mass along the phase direction is then given by
\begin{equation}
-\frac{|a_H|H^3}{4}{M_*}\left(-\frac{1}{X_0^2}-\frac{16}{\phi_0^2}\right)
\simeq\frac{4|a_H|H^3M_*}{\phi_0^2}
=4|a_H|H^2,
\end{equation}
where we have used the fact that $\phi_0\ll X_0$.
The mass along the phase direction is of order $H$, so the phase condensate rapidly rolls down to its minimum.
Since the phase minimum of this Hubble-induced $A$ term differs from the soft SUSY breaking-induced $A$ term,
the AD field obtains angular momentum in the complex plane, generating the lepton number.
Note that we have one more phase direction orthogonal to the above massive direction.
This does not appear in the potential, so is a massless mode corresponding to the axion of the spontaneously broken PQ symmetry.
In the limit of $\phi_0\ll X_0$, the massive mode is mostly $a_\phi$-like 
and the massless mode is mostly $a_X$-like.

\subsection{Baryon asymmetry along $LH_u$ direction}

Now let us evaluate the lepton number generated through the AD mechanism.
The massive phase mode automatically cancels the imaginary part of the Hubble induced $A$-term potential, so it does not significantly contribute to lepton number generation.
Thus, as in ordinary AD leptogenesis, 
lepton number is determined by the ordinary $A$-term which depends on 
the hidden sector SUSY breaking ({\it i.e.}, gravitino mass in gravity mediation).
The lepton number obeys the equation,\footnote{
	While the lepton number is violated by the $(LH_u)^2$ term in the superpotential after integrating out $N$,
	the PQ number is exactly conserved (except for the small instanton effect).
}
\begin{equation}
\dot{n}_L+3Hn_L\simeq\frac{m_{\rm soft}}{2M_*}{\rm Im}(a_m\phi^4).
\end{equation}
The baryon number to entropy ratio is obtained as~\cite{Fujii:2001zr}
\begin{equation}
\frac{n_B}{s}\simeq0.029\frac{M_*T_R}{M_P^2}\left(\frac{m_{\rm soft}|a_m|}{H_{\rm osc}}\right)\delta_{\rm ph},
\label{eq:bary-to-ent_wo_dilut}
\end{equation}
where $\delta_{\rm ph}$ represents an effective CP violating phase, and  
$H_{\rm osc}$ is the Hubble parameter when the $\phi$ field starts to oscillate. 
Taking account of thermal effects on the AD potential~\cite{Allahverdi:2000zd,Anisimov:2000wx}, the latter is determined by
\begin{equation}
H_{\rm osc}\simeq\max\left[
m_{\phi}, ~H_i, ~\alpha_ST_R\left(\frac{a_gM_P}{M_*}\right)^{1/2}
\right],
\label{eq:Hosc}
\end{equation}
where
\begin{equation}
H_i=\min\left[
\frac{1}{f_i^4}\frac{M_PT_R^2}{M_*^2},~(c_i^2f_i^4M_PT_R^2)^{1/3}
\right].
\end{equation}
Here, the $f_i$ are coupling constants of $\phi$ with particles in thermal background, $c_i$ and $a_g(=1.125)$ are real positive constants of order unity, and $T_R$ is the reheating temperature after inflation.
From Eq.~(\ref{eq:Hosc}), $H_{\rm osc}$ is nearly $m_{\phi}$ when the temperature is small.
In the case of a large reheat temperature, the $\phi$ oscillation can commence earlier due to thermal effects which result in the second and third terms inside the bracket of Eq.~(\ref{eq:Hosc}).
The detailed physical aspects are explained in Ref.~\cite{Hamaguchi:2002vc} and references therein.
If $m_{\phi}=10$ TeV and $M_{*}=7.2\times10^{23}$ GeV, such early oscillation occurs for $T_R\gtrsim10^7$ GeV.

For an illustration, we show a formula for a case where the early oscillation does not occur.
In such a case, $H_{\rm osc}=m_{\phi}$, so the baryon-number-to-entropy ratio becomes
\begin{eqnarray}
\frac{n_B}{s}\simeq 3.6\times10^{-8} \delta_{\rm ph} 
\left(\frac{T_R}{10^7\mbox{ GeV}}\right)
\left(\frac{10^{-4}\mbox{ eV}}{m_{\nu1}}\right)
\left(\frac{10^{12}\mbox{ GeV}}{f}\right)
\left(\frac{X_0}{M_P}\right),
\end{eqnarray}
where we also assume $m_{\phi}=|a_m| m_{\rm soft}$.
In this scenario, however, saxion will dominate the universe, and its decay produces entropy dilution.
In order to obtain the final baryon asymmetry after saxion decay, the entropy dilution must be taken into account.
We will consider the entropy production in the following subsection.

Before closing this subsection, let us comment on the possible lepton number violation during the saxion oscillation. Since the field value of $X$ can become small during its oscillation, the effect of the lepton number violation, induced by the effective superpotential \eqref{eq:ADeff} or the corresponding $A$-term, may become large. Here, as shown in Sec.~\ref{sec:PQ_dyn}, the $\mu$-term interaction in Eq.~\eqref{eq:lagPQ},
\begin{equation}
W_{\mu}=\frac{g_{\mu}Y^2}{M_P}H_uH_d\,,
\label{eq:mu}
\end{equation}
plays an important role. Assuming $m_X\ll m_\phi$, the lepton number violation during the saxion oscillation is small enough to maintain the generated lepton asymmetry by the AD mechanism.
As we shall see in the next subsection, the $\mu$-term interaction~\eqref{eq:mu} also plays a key role to determine the saxion decay, and hence the final baryon asymmetry.

\subsection{Saxion decay in DFSZ model}
\label{sec:sax_dec}

We have discussed how the dynamical PQ breaking scale can enhance the 
baryon asymmetry.
For the final result, one crucial point to consider is the entropy production from saxion decay.
In the DFSZ model, saxion interactions with the standard model particles and their superpartners are realized in the $\mu$-term interaction~\eqref{eq:mu}.
Once $X$ and $Y$ settle down to the current value of the PQ symmetry breaking scale, $X\sim Y\sim f$, 
this superpotential generates the $\mu$-term, 
\begin{equation}
\mu\sim \frac{g_{\mu}f^2}{M_P},
\end{equation}
and also interactions between the axion superfield and the 
Higgs supermultiplets.

Through this interaction, the saxion dominantly decays into Higgsino states 
if they are kinematically allowed.
Its decay rate is approximately given by~\cite{Bae:2013hma}\footnote{
In the numerical calculation in Sec.\ref{sec:susy_scale}, we use the saxion decay rates including phase space and mixings in~\cite{Bae:2013hma}.
}
\begin{equation}
\Gamma(\sigma\to 2\widetilde{H})\simeq\frac{1}{4\pi}\left(\frac{\mu}{f_a}\right)^2m_{\sigma}.
\label{eq:saxdec1}
\end{equation}
Note that we have used here $f_a=2f$ under assumption of $X=Y=f$ in the 
present universe, so quantities related to axion dark matter is determined by $f_a/N_{\rm DW}$ ($N_{\rm DW}$: domain wall number) as the usual normalization.
The decay temperature is
\begin{equation}
T_{\sigma}^{(m_{\sigma}>2\mu)}\simeq25\mbox{ GeV}\left(\frac{90}{g_*}\right)^{1/4}
\left(\frac{\mu}{\mbox{TeV}}\right)\left(\frac{10^{12}\mbox{ GeV}}{f_a}\right)
\left(\frac{m_{\sigma}}{10\mbox{ TeV}}\right)^{1/2}.
\end{equation}
If saxion decays into Higgsino states are disallowed, it dominantly decays into the light Higgs and gauge bosons.
The decay rate in such a case is given by
\begin{equation}
\Gamma(\sigma\to hh,W^+W^-,ZZ)\simeq\frac{2}{\pi}\frac{\mu^4}{f_a^2}\frac{1}{m_{\sigma}},
\label{eq:saxdec2}
\end{equation}
and the decay temperature becomes
\begin{equation}
T_{\sigma}^{(m_{\sigma}<2\mu)}\simeq70\mbox{ GeV}
\left(\frac{90}{g_*}\right)^{1/4}
\left(\frac{\mu}{\mbox{TeV}}\right)^2\left(\frac{10^{12}\mbox{ GeV}}{f_a}\right)
\left(\frac{100\mbox{ GeV}}{m_{\sigma}}\right)^{1/2} .
\end{equation}
From the above decay temperature for each case, one finds the entropy dilution factor
\begin{equation}
\Delta=\max\left[\frac18T_R\left(\frac{X_0}{M_P}\right)^2\frac{4}{3T_{\sigma}},1\right].
\label{eq:dilut}
\end{equation}
Here we have included the case where $T_R$ is small so that the 
saxion decays before the reheating process is over.
The final baryon asymmetry is determined by the amount of asymmetry when the AD mechanism completes, Eq.~(\ref{eq:bary-to-ent_wo_dilut}) and by the 
dilution factor, Eq.~(\ref{eq:dilut}):
\begin{equation}
\left(\frac{n_B}{s}\right)_{\rm final}
=0.029\frac{M_*T_R}{M_P^2}\left(\frac{m_{\rm soft}|a_m|}{H_{\rm osc}}\right)\delta_{\rm ph}
\times\frac{1}{\Delta}.
\label{eq:nBs_final}
\end{equation}
For the case where saxion dominantly decays into Higgsinos,
\begin{equation}
\frac{n_B}{s}=1.1\times10^{-12}~\delta_{\rm ph}
\left(\frac{10^{-4}\mbox{ eV}}{m_{\nu1}}\right)
\left(\frac{10^{12}\mbox{ GeV}}{f_a}\right)^2
\left(\frac{X_0}{M_P}\right)^{-1}
\left(\frac{90}{g_*}\right)^{1/4}
\left(\frac{\mu}{\mbox{TeV}}\right)
\left(\frac{m_{\sigma}}{10\mbox{ TeV}}\right)^{1/2},
\label{eq:bau_large}
\end{equation}
or for the case where saxion dominantly decays into light Higgs and gauge bosons,
\begin{equation}
\frac{n_B}{s}=3.0\times10^{-12}~\delta_{\rm ph}
\left(\frac{10^{-4}\mbox{ eV}}{m_{\nu1}}\right)
\left(\frac{10^{12}\mbox{ GeV}}{f_a}\right)^2
\left(\frac{X_0}{M_P}\right)^{-1}
\left(\frac{90}{g_*}\right)^{1/4}
\left(\frac{\mu}{\mbox{TeV}}\right)^2
\left(\frac{100\mbox{ GeV}}{m_{\sigma}}\right)^{1/2} .
\label{eq:bau_small}
\end{equation}
From this it is easily seen that the observed baryon-number-to-entropy-ratio can be obtained
for relatively large neutrino mass $m_{\nu1}=10^{-4}$ eV if the 
PQ breaking scale is near the Planck scale in the beginning and settles 
to $10^{11}$ GeV at the present universe.

We will see numerical results for some example parameter regions in Sec.~\ref{sec:susy_scale}.

\section{Dynamics of PQ breaking fields}
\label{sec:PQ_dyn}

In this section, we discuss the dynamics of the PQ breaking fields in 
order to investigate the realization of the Planck scale PQ breaking in 
the early stage and lepton number conservation at the late stage.


\subsection{PQ breaking at the Planck scale}
Let us first examine the scalar potential of $X$ and $Y$ for large $H$ to check if the PQ scale is ${\cal O}(M_P)$.
%
%
We have to consider the supergravity potential which is given by
\begin{equation}
V=e^{K/M_P^2}\left(D_iW K^{i\bar{j}}D_{\bar{j}}W^*-\frac{3}{M_P^2}|W|^2\right),
\end{equation}
where $D_iW=W_i+K_iW/M_P^2$.
We assume that the effect of the AD field $\phi$ is negligible.
The K\"ahler potential and superpotential are given by
\begin{eqnarray}
K&=&|X|^2+|Y|^2+|Z|^2+|I|^2+\frac{b}{M_P^2}|I|^2|X|^2,\\
W&=&\eta Z(XY-f^2),
\end{eqnarray}
where $I$ is the inflaton field.
Note that only $X$ has non-minimal coupling with the inflaton in $K$.
If $b>1$, one can obtain a negative Hubble-induced mass term for $X$ 
and thus $X$ develops a large VEV.
Let us see the scalar potential in detail.

In this discussion, $Z$ obtains a mass of $|X|$ and $|Y|$ ($> f$), so its VEV is zero up to of the order of the gravitino mass: ${\cal O}(m_{3/2})$. 
Thus we can safely neglect the dynamics of $Z$.
If $I\ll M_P$ during inflation, the inflaton energy is dominantly determined by the $F$-term potential, {\it i.e.} $K_I\ll M_P$, $|W|/M_P\ll W_I$ and $D_IW\simeq W_I\simeq F_I$.
One can simplify the inflaton potential: $V(I)\sim |F_I|^2\sim H^2M_P^2$.
In these circumstances, the scalar potential of $X$ and $Y$ is given by
\begin{equation}
V=e^{(|X|^2+|Y|^2)/M_P^2}\left(\eta^2|XY-f^2|^2+\frac{|F_I|^2}{1+b|X|^2/M_P^2}\right), \label{VXY}
\end{equation}
Let us define $\langle X\rangle=x$ and $\langle Y\rangle=y$.
The extremum condition is obtained as
\begin{eqnarray}
0=\frac{\partial V}{\partial x}&=&
\left[
\eta^2(xy-f^2)\left(\frac{2x}{M_P^2}(xy-f^2)+2y\right)
-\frac{|F_I|^2}{1+bx^2/M_P^2}\left(\frac{2x}{M_P^2}-\frac{2bx}{M_P^2}\frac{1}{1+bx^2/M_P^2}\right)
\right]\nonumber\\
&&\times e^{(x^2+y^2)/M_P^2},\label{eq:dVdx}\\
0=\frac{\partial V}{\partial y}&=&
\eta^2(xy-f^2)\left(\frac{y}{M_P^2}(xy-f^2)+2x\right)e^{(x^2+y^2)/M_P^2}.
\end{eqnarray}
From the second equation, we find that 
\begin{equation}
xy=f^2\qquad\mbox{or}\qquad xy^2-f^2y+M_P^2x=0.
\end{equation}
Since the second solution leads to a trivial solution $x=y=0$ 
we select the first solution.
From Eq.~(\ref{eq:dVdx}), we obtain the solution for $x$,
\begin{equation}
x=\left(1-\frac{1}{b}\right)^{1/2}M_P.
\end{equation}
For $b>1$, $x$ develops an ${\cal O}(M_P)$ VEV as long as $|F_I|^2/M_P^2\gg m_{3/2}^2$.
It is also evident that $X$ obtains a (negative) mass squared of the order of $H^2$ during the inflaton domination.


\subsection{Saxion oscillation and lepton number conservation}

As argued in the previous subsection, $X$ stays at $X\sim M_P$ until the Hubble parameter drops down to $m_{3/2}$.
After that, the saxion begins a coherent oscillation around the minimum $x\sim y \sim f$ with an initial amplitude of $\sim M_P$.
Since the scalar field orthogonal to the $F$-flat direction (saxion) has a mass of $\sim f$, which is much higher than the soft mass scale,
we can safely set $XY = f^2$ to integrate out either $X$ or $Y$.
Then the scalar potential along the $F$-flat direction $XY=f^2$ and the AD field $\phi$ reads
\begin{align}
	V \simeq m_X^2 |X|^2  + m_\phi^2|\phi|^2 +m_Y^2 \frac{f^4}{|X|^2} + \frac{g_\mu^2 f^8}{M_P^2} \left| \frac{\phi}{X^2} \right|^2,  \label{VXAD}
\end{align}
where $m_X$ and $m_Y$ are the soft SUSY breaking mass of $X$ and $Y$, respectively.
The last term comes from $W_\mu$ (\ref{eq:mu}).
The third and fourth terms act as the effective potential for $X$ and they 
prevent $X$ from being very small during the oscillation.
Let us denote by $X_{\rm max}$ the maximum value of $X$ during each $X$ oscillation, which adiabatically becomes smaller due to the Hubble expansion $(X_{\rm max} \propto a^{-3/2})$.
Then we can define $X_{\rm min}$, the minimum value of $X$ during each oscillation.
For a large AD field value $\phi$, the last term is important to determine $X_{\rm min}$.
Thus we can evaluate $X_{\rm min}$ as
\begin{align}
	X_{\rm min}\sim
	{\rm max} \left[
	 \frac{f^2}{X_{\rm max}} \left(\frac{g_\mu X_{\rm max} |\phi|}{M_P m_X} \right)^{1/2},~~~
	 \frac{m_Y}{m_X}\frac{f^2}{X_{\rm max}}
	 \right].
	 \label{eq:Xmin}
\end{align}
Since $X_{\rm max} \sim M_P$ and $|\phi| \gg m_X$ just after the saxion oscillation, 
$X_{\rm min}$ is generically much larger than the soft mass scale, meaning that the RHN masses cannot be as small as the soft mass during 
saxion oscillation and hence the procedure to integrate out the 
RHN to obtain the effective potential of the AD field is justified.

Now let us consider the lepton number violation after the 
$X$ begins to oscillate. The lepton number follows:
\begin{align}
	\dot n_L + 3Hn_L = \frac{y_\nu^2 m_{\rm soft}}{\lambda X}{\rm Im}(a_m \phi^4).
\end{align}
As discussed above, $X$ oscillates between $X_{\rm max}$ and $X_{\rm min}$ in a time scale $m_X^{-1}$ where $X_{\rm min}$ is given by Eq.~\eqref{eq:Xmin}.
The most dangerous $L$ violation may happen around $X\sim X_{\rm min}$ at which the $L$-violating operator becomes large.
The time interval $\Delta t$ during which $X \sim X_{\rm min}$ is estimated 
from the equation of motion
\begin{align}
	\ddot X \sim \frac{\mu^2 f^4 \phi^2}{X^5} ~~~\to~~~ X_{\rm min}\sim \ddot X (\Delta t)^2~~~\to~~~\Delta t \sim \frac{X_{\rm min}^3}{\mu f^2 \phi}.
\end{align}
During this time interval, the $L$ number changes as
\begin{align}
	\Delta n_L \sim \frac{y_\nu^2 m_{\rm soft} \phi^4}{\lambda X}\Delta t \sim \frac{y_\nu^2}{\lambda} \frac{g_\mu m_{\rm soft} \phi^4 f^2}{\mu m_X M_P X_{\rm max}}.
\end{align}
Using $n_L \sim m_{\rm soft}\phi^2$, we obtain
\begin{align}
	\frac{\Delta n_L}{n_L} \sim \frac{y_\nu^2}{\lambda} \frac{\phi^2}{m_X X_{\rm max}}.
\end{align}
Since $\phi^2$ in the numerator decreases faster than $X_{\rm max}$ in the denominator,
this takes a maximum value just after the $X$ begins to oscillate $H \sim m_X$.
\begin{align}
	\left(\frac{\Delta n_L}{n_L} \right)_{H \sim m_X} \sim \frac{y_\nu^2}{\lambda}  \frac{\phi_{H=m_X}^2}{m_X M_P}.
\end{align}
This must be smaller than 1 to ensure the conservation of lepton number.
If thermal effects are neglected and $m_X \ll m_\phi$, we have $\phi_{H=m_X}^2 \sim m_\phi M_* (m_X/m_\phi)^2$ and it becomes
\begin{align}
	\left(\frac{\Delta n_L}{n_L} \right)_{H \sim m_X} \sim \frac{m_X}{m_\phi} \ll 1\,.
\end{align}
Thus, the lepton number violation during the saxion oscillation can be neglected as far as $m_X \ll m_\phi$ is satisfied.



\section{Baryon number and SUSY scale}
\label{sec:susy_scale}


We have discussed how PQ symmetry breaking accommodates the 
baryon asymmetry with a sizable neutrino mass when the PQ scale varies during and after inflation.
In this scenario, the entropy dilution from saxion decay indeed plays a substantial role for determining the final value of the baryon asymmetry.
The saxion decay rate depends on its mass and the 
$\mu$-term as shown in Eqs.~(\ref{eq:saxdec1}) and (\ref{eq:saxdec2}).
In many cases, the saxion mass and $\mu$-term are related to the 
soft SUSY breaking scale. 
In particular, $\mu$-term is a measure of fine-tuning of the electroweak symmetry breaking.
Therefore, it leads us to discuss the soft SUSY scale and fine-tuning from the measured baryon asymmetry.

Since the saxion is linked to the axion which is the Nambu-Goldstone boson of broken PQ symmetry,
it is massless in the supersymmetric limit.
When SUSY is broken, however, the saxion (and also the axino) acquires a mass.
The saxion mass is typically of the gravitino mass order although it can be either larger or smaller than the gravitino mass in some models~\cite{Goto:1991gq,Chun:1992zk,Chun:1995hc,Bae:2014efa}. 
On the other hand, as shown in the Sec.~\ref{sec:PQ_dyn}\,B,  the 
saxion mass ({\it i.e.} $m_X$) is required to be smaller than the AD field mass in order to not spoil lepton number generation.
In this regard, we consider a rather small saxion mass compared to the 
AD field mass, {\it i.e.} $m_{\sigma}\sim m_X\lesssim m_{\phi}/10$.


In contrast to the saxion mass, the $\mu$-term is a supersymmetric parameter,
so its origin can be different from the SUSY breaking.
In models with PQ symmetry breaking, the $\mu$-term can be generated from PQ symmetry breaking
through non-renormalizable interactions~\cite{Kim:1983dt}.\footnote{In models with radiative PQ symmetry breaking~\cite{Murayama:1992dj,Gherghetta:1995jx,Choi:1996vz,Abe:2001cg,Nakayama:2012zc,Bae:2014yta}, 
SUSY breaking leads to PQ symmetry breaking via RG running, 
and thus the $\mu$-term generated by PQ symmetry breaking is related to the soft SUSY scale.
In this work, however, we are agnostic as to the origin of the PQ scale $f$ in Eq.~(\ref{eq:lagPQ}).}
In Sec.~\ref{sec:sax_dec}, we have discussed the $\mu$-term generation via an
interaction suppressed by the Planck scale as shown in Eq.~(\ref{eq:mu}).
In such a case, the $\mu$-term is typically ${\cal O}(f_a^2/M_P)$.
However, the suppression scale for this interaction can be different from the Planck scale ({\it e.g.} the grand unification scale), while the coupling constant ($g_\mu$) for this interaction can be smaller than unity.
For this reason, we will consider $\mu$ as an independent parameter of the model in the following discussions.

In order to achieve successful electroweak symmetry breaking, the soft SUSY breaking scale and $\mu$-term must coincide with each other since they need to satisfy the relation,
\begin{equation}
\frac{m_Z^2}{2}\simeq -m_{H_u}^2-\mu^2,
\end{equation}
where $m_{H_u}$ is the soft mass term for the up-type Higgs at the weak scale.
For a natural model, these three quantities above need to be comparable 
to one another so that no dramatic cancellation takes place. 
If $m_{H_u}$ and $\mu$ are much larger than $m_Z$, on the other hand, 
fine-tuning arises.
Although it is hard to quantify the level of fine-tuning without specifying the whole SUSY spectrum, 
we can roughly see how much fine-tuning is required from the size of $\mu$-term (or equivalently $m_{H_u}$)~\cite{Baer:2012up}:
\begin{equation}
\Delta_{\rm EW} \sim \frac{\mu^2}{m_Z^2/2}.
\end{equation}
The baryon asymmetry depends on the soft SUSY breaking scale 
when the AD mechanism works as described in Eq.~(\ref{eq:bary-to-ent_wo_dilut}).
It is also dependent on the saxion decay rate which is 
determined by the saxion mass (soft SUSY scale), the $\mu$-term and the 
PQ breaking scale as shown in Eqs.~(\ref{eq:bau_large}) and (\ref{eq:bau_small}).
Therefore, by requiring $n_B/s\simeq 10^{-10}$, we can obtain {\it the relation between the lightest neutrino mass and $\mu$-term}.

\begin{figure}
\includegraphics[width=8cm]{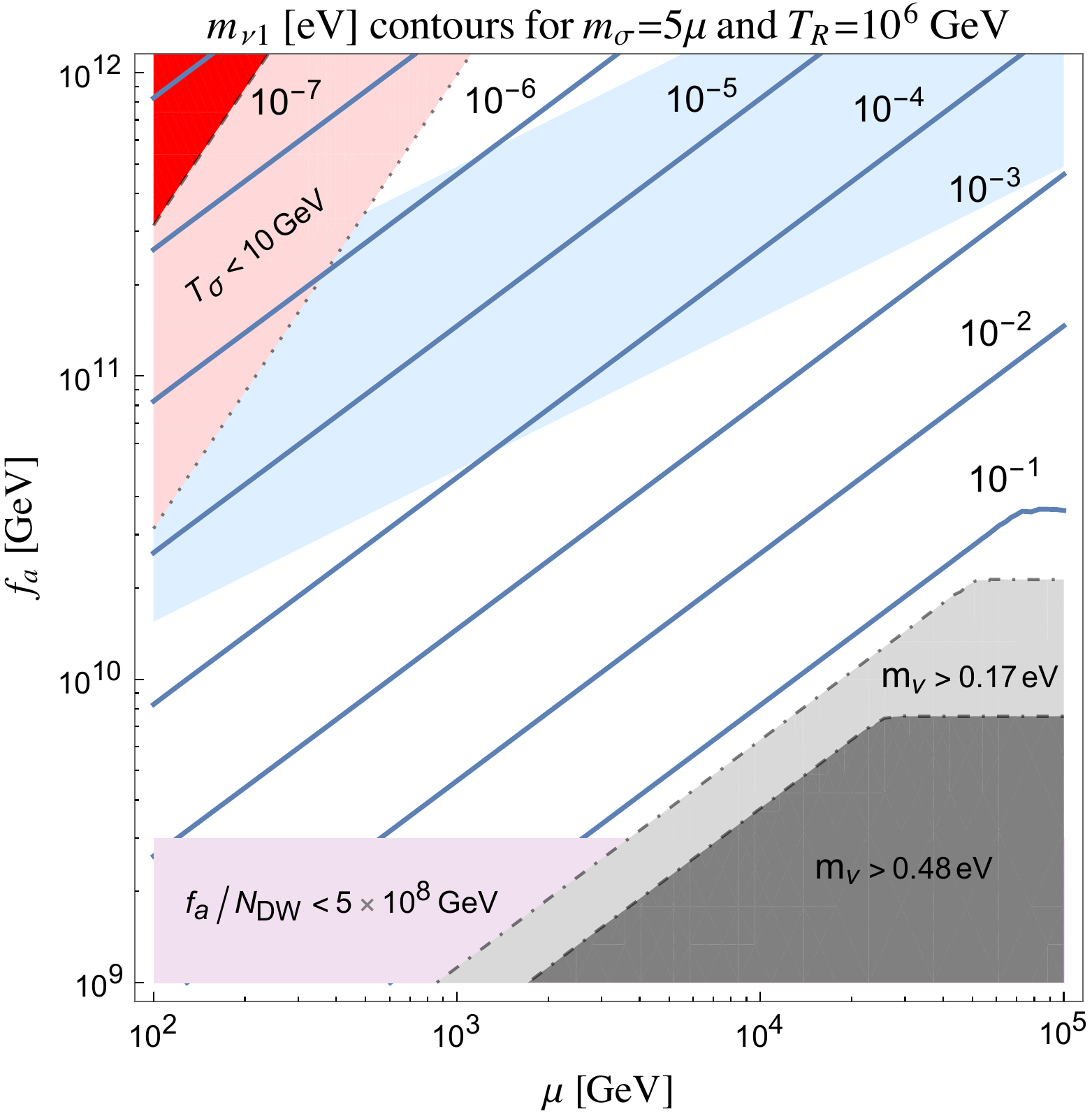}\quad
\includegraphics[width=8cm]{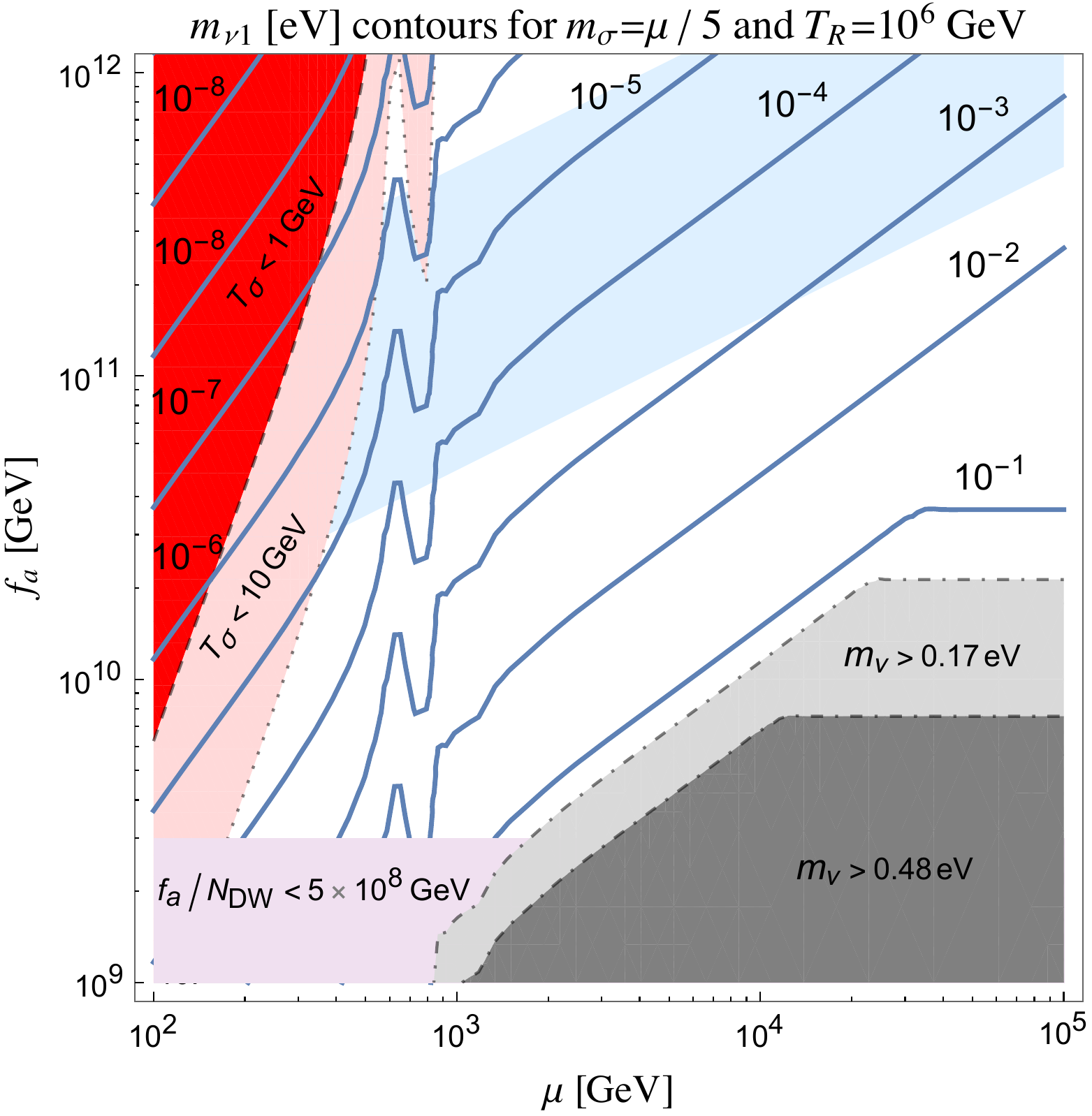}
\caption{
Contours of $m_{\nu 1}$\,[eV] on the $(\mu,f_a)$ plane to reproduce the observed baryon asymmetry.
In the left (right) panel we have taken $m_\sigma = 5\mu$ $(m_\sigma=\mu/5)$ while
$m_{\rm soft}=m_{\phi}=10m_{\sigma}$ ($m_{\rm soft}=m_{\phi}=50m_{\sigma}$). 
The light (dark) red shaded region corresponds to $T_\sigma < 10$\,GeV (1\,GeV) and the grey shaded region is excluded by the
KamLAND-Zen experiment.
The light-gray shaded region is constrained by Planck+BAO~\cite{Ade:2015xua}.
The light-purple shaded region indicates bound from SN1987A~\cite{Kim:2008hd}.
The blue shaded region corresponds to $\mu = (0.01-1)f_a^2/M_P$.
}
\label{fig:fig1}
\end{figure}

Fig.~\ref{fig:fig1} shows illustrative contours of neutrino masses which produce the desired baryon asymmetry, $n_B/s=10^{-10}$ for given 
values of $f_a$ and $\mu$.
In the left panel, we take $m_{\sigma}=5\mu$ for which  saxion decays into Higgsino states are allowed while in the right panel, we take $m_{\sigma}=\mu/5$ for which saxion can decay only into SM particles.
In order to maintain the generated lepton asymmetry, 
$m_{\rm soft}=m_{\phi}=10m_{\sigma}$ and $m_{\rm soft}=m_{\phi}=50m_{\sigma}$ are taken respectively for each case.
The gray shaded region shows parameter space where the lightest neutrino mass is larger than the
KamLAND-Zen bound~\cite{KamLAND-Zen:2016pfg}.
We also show a bound from Planck+BAO constraint on the sum of neutrino masses, $(\sum m_{\nu})<0.17$ eV~\cite{Ade:2015xua}.
The light-purple shaded region shows the bound from SN1987A~\cite{Kim:2008hd}.
The (light-)red shaded region shows parameter space where the saxion decay temperature is 
smaller than 1 GeV (10 GeV).
The blue shade indicates the region for which $\mu=(0.01-1)f_a^2/M_P$.
We consider fixed $T_R=10^6$ GeV 
since larger $T_R$ does not change or does suppress $n_B/s$ (see Eq.~(\ref{eq:nBs_final})).
In the case where $n_B/s$ is suppressed, 
it requires a smaller neutrino mass that is less attractive.

From the figure, it is clearly shown that neutrino mass is large for large $\mu$ and small $f_a$ 
while it becomes smaller for small $\mu$ and large $f_a$.
This feature stems from the saxion decay temperature.
The saxion decay temperature is enhanced by the $\mu$-term while suppressed by $f_a$.
It is also of great importance that small $f_a$ is good for obtaining a 
flatter direction during lepton number generation as shown in Eq.~(\ref{eq:M_*}).
%
For $\mu\gtrsim1$ TeV and $f_a\lesssim10^{10}$ GeV, 
our model predicts a rather large neutrino mass so that it is constrained by recent neutrinoless 
double beta decay ($0\nu\beta\beta$) experiment.
For this constraint, we take a conservative bound from KamLAND-Zen, $m_{\nu} < 0.48$ eV~\cite{KamLAND-Zen:2016pfg}.
From the lower-right corner ($\mu\sim 10^5$ GeV, $f_a\sim 10^{10}$ GeV) 
to the upper-left corner ($\mu\sim 10^2$ GeV, $f_a\sim 10^{12}$ GeV), 
the resulting neutrino mass scans over $10^{-1}-10^{-8}$ eV.

For large $f_a\sim10^{12}$ GeV, in order to obtain a natural value of the lightest neutrino mass, $\sim 10^{-4}$ eV
in cases of both $m_{\sigma}=5\mu$ and $m_{\sigma}=\mu/5$,
the $\mu$-term is required to be tens of TeV.
In such cases, the fine-tuning in the electroweak symmetry breaking is of permyriad ($10^{-4}$) order.
For a smaller PQ scale, $f_a\sim 10^{10}$ GeV, $\mu$ can be a few hundred GeV to achieve $m_{\nu1}\sim 10^{-4}$ eV, so the model is much less fine-tuned.
If we constrain the $\mu$-term to make fine-tuning better than a 
percent level, {\it i.e.} $100$ GeV$\lesssim \mu\lesssim300$ GeV, $10^{-6}$ eV$\lesssim m_{\nu1}\lesssim10^{-4}$ eV can be achieved for $10^{10}$ GeV$\lesssim f_a\lesssim10^{11}$ GeV.
This region is well-matched with the parameter space where $\mu$ can be determined by the Planck suppressed interaction, Eq.~(\ref{eq:mu}).
Moreover, it is good for the {\it mixed axion-higgsino dark matter scenario}~\cite{Bae:2013hma,Bae:2014rfa}.
 
Before closing this section, it is worth noting that the saxion decay temperature can be smaller than $1-10$ GeV for small $\mu$ and large $f_a$ as indicated by red shaded regions in Fig.~\ref{fig:fig1}.
The region with $T_{\sigma}\lesssim10$ GeV may cause the saxion to 
decay after freeze-out of weakly interacting massive particle (WIMP) 
dark matter and thus such late decay affects the 
WIMP dark matter density due to entropy production and/or non-thermal dark matter production,
although it is not possible to make a concrete analysis without a specific SUSY spectrum.
Moreover, for the region with $T_{\sigma}\lesssim1$ GeV, 
the saxion decays after coherent oscillation of axion commences, so it 
affects the axion dark matter density, too.

\section{Cosmological implications of Peccei-Quinn sector}  \label{sec:PQ}

In this section we discuss some cosmological implications related 
to the PQ sector: axion isocurvature perturbation and axino production.

\subsection{Axion isocurvature perturbation}

There is a (nearly) massless Goldstone boson from the spontaneous breakdown of the U(1)$_{\rm PQ}$ symmetry, the axion,
which can also be interpreted as a massless majoron in our model~\cite{Langacker:1986rj}.
Such a massless boson potentially causes several cosmological problems~\cite{Kawasaki:2013ae}.

In our setup, the PQ symmetry is already broken during inflation and is not restored thereafter.
Thus there is no axionic domain wall problem.\footnote{
	Domain wall formation in a SUSY axion model similar to the present one was discussed in Ref.~\cite{Kasuya:1996ns},
	although it is not clear what is the necessary condition for the domain wall formation.
}
The PQ scalar $X$ obtains a large VEV of $\sim M_P$ in our setup, hence the PQ scale during inflation is
much higher than that in the present universe.
It significantly suppresses the axion isocurvature perturbation~\cite{Linde:1990yj,Linde:1991km,Kawasaki:1995ta,Kawasaki:2008jc}.
Since the massless axion mode almost consists of the phase component of $X$ for $|X|\gg f$,
the effective PQ scale during inflation is simply given by $|X| = X_{\rm inf}\sim M_P$.\footnote{
	If $H_u$ or $H_d$ has a larger field value than $X$, the effective PQ scale is given by $|H_u|$ or $|H_d|$~\cite{Choi:2015zra,Nakayama:2015pba}.
	This is not the case in our model.
}

The magnitude of CDM isocurvature perturbation is then given by
\begin{align}
	\mathcal P_{S_{\rm CDM}} \simeq r^2 \left(\frac{H_{\rm inf}}{\pi X_{\rm inf}\theta_a}\right)^2.
\end{align}
where $H_{\rm inf}$ denotes the Hubble scale during inflation,
$\theta_a$ denotes the initial misalignment angle of the axion and
$r$ denotes the fraction of present axion energy density in the matter energy density: $r\equiv (\Omega_a h^2)/(\Omega_m h^2)$.
The final axion density is given by~\cite{Oh2_axion}
\begin{align}
	\Omega_a h^2 \simeq 0.18\,\theta_a^2 \left( \frac{f_a/N_{\rm DW}}{10^{12}\,{\rm GeV}} \right)^{1.19}. \label{Omega_a}
\end{align}
Here we have assumed that there is no dilution of the axion density due to the saxion decay.
The Planck constraint on the uncorrelated isocurvature perturbation \cite{Ade:2015lrj} reads 
\begin{align}
	H_{\rm inf} \lesssim 7\times 10^{13}\,{\rm GeV}\,\theta_a^{-1}\left( \frac{10^{12}\,{\rm GeV}}{f_a/N_{\rm DW}} \right)^{1.19} \left( \frac{X_{\rm inf}}{M_P} \right).
\end{align}
This constraint is easily satisfied for most inflation models
since $X_{\rm inf}=X_0\sim M_P$ in our scenario.

\subsection{Axino production}

The axino is the fermionic superpartner of the axion, 
consisting of the fermionic components of $X$ and $Y$ with a small mixture of higgsino.
It obtains a mass of $m_{\tilde a} = \eta\left<Z\right>\simeq m_{3/2}$.
It has a relatively long lifetime if it is not the lightest SUSY particle (LSP)~\cite{Chun:2011zd,Bae:2011jb,Bae:2011iw}.
Its decay width is approximately given by
\begin{align}
	\Gamma_{\tilde a} \sim \frac{2}{\pi}\frac{\mu^2 m_{\tilde a}}{f_a^2},
\end{align}
which is comparable to the saxion.
Thus the axino can have significant impacts on cosmology.

The dominant axino production process is the thermal one.\footnote{
	The direct saxion decay into the axino pair can be kinematically forbidden for $m_\sigma < 2 m_{\tilde a}$.
}
The axino thermal production in the DFSZ model comes from the combination of higgsino decay/inverse decay, scatterings of Higgs and weak gauge bosons and also top/stop scatterings~\cite{Chun:2011zd,Bae:2011jb,Bae:2011iw}.
There it was found that the production is dominated at $T \sim m_{\rm soft}$ in general and the abundance is independent of the 
reheating temperature $T_R$ as long as $T_R \gg m_{\rm soft}$.

In our case, the saxion decay temperature $T_\sigma$ can be lower than $m_{\rm soft}$ and hence there is a dilution of the preexisting axino abundance.
First, let us consider the case $T_\sigma > m_{\rm soft}$.
One of the main contributions may be the heavy Higgs decay into the higgsino plus axino with the partial decay rate $\Gamma_H \sim \mu^2 m_H/(4\pi f_a^2)$ where the heavy Higgs mass is assumed to be $m_H \sim m_{\rm soft}$. The axino abundance is then estimated as
\begin{align}
	Y_{\tilde a} \sim  \left(\frac{n_H}{s} \frac{\Gamma_H}{H}\right)_{T\sim m_{\rm soft}} \sim 10^{-7}\left( \frac{10^{12}\,{\rm GeV}}{f_a} \right)^2.
\end{align}
In the opposite case $T_\sigma < m_{\rm soft}$, we must take the dilution factor into account.
Noting that $T^4 \sim T_\sigma^2 H M_P\propto a^{-3/2}$ during 
saxion domination, the resultant axino abundance is given by
\begin{align}
	Y_{\tilde a} \sim  \left(n_H \frac{\Gamma_H}{H}\right)_{T\sim m_{\rm soft}} \left( \frac{a(m_{\rm soft})}{a(T_\sigma)} \right)^3\frac{1}{s(T_\sigma)}
	\sim 10^{-7}\left( \frac{10^{12}\,{\rm GeV}}{f_a} \right)^2\left( \frac{T_\sigma}{m_{\rm soft}} \right)^7.
\end{align}
Due to the dilution, the axino abundance can be suppressed.
When the axino decays into LSP ({\it e.g.} neutralinos), the LSP density is determined by its re-annihilation rate~\cite{Choi:2008zq},
\begin{equation}
Y_{\rm LSP}\sim\left(\frac{H}{\langle \sigma v\rangle s}\right)_{T\sim T_{\tilde{a}}},
\end{equation}
where $T_{\tilde{a}}$ is the axino decay temperature.
Thus the dark matter density from axino decay highly depends on 
details of the axino decay as well as upon 
its annihilation rate at $T_{\tilde{a}}$.

\section{Conclusions}  \label{sec:conc}

In this paper we have reconsidered the AD leptogenesis in a scenario where 
the RHN mass is dynamical.
If the RHN mass is generated by the PQ field, it naturally takes hierarchically different values between the early universe and the present epoch.
In particular, the PQ scalar can be stabilized at the Planck scale in the early universe until the lepton asymmetry is generated,
which makes leptogenesis much more efficient than in the ordinary scenario.
The predicted lightest neutrino mass to reproduce the observed baryon asymmetry can be close to the neutrino mass differences
known from the neutrino oscillation data.
It significantly relaxes the problem of the ordinary AD leptogenesis scenario in which the lightest neutrino mass should be hierarchically smaller
than the other two neutrinos.

In order to realize this scenario, we have considered the DFSZ model 
which provide a solution to the strong $CP$ problem and generates the 
$\mu$-term and the RHN mass. 
Since the final baryon asymmetry depends on the saxion decay, it is related to the $\mu$-term and to the electroweak fine-tuning.
As a result, a comparatively large mass of the lightest neutrino, 
$10^{-6}-10^{-4}$ eV, is predicted 
for $10^{10}$ GeV$\lesssim f_a\lesssim10^{11}$ GeV and $100$ GeV$\lesssim \mu\lesssim300$ GeV.
In this model, therefore, it is possible to accommodate successful AD leptogenesis with a natural neutrino mass and natural electroweak symmetry breaking.
In addition, due to the very large PQ scale during inflation, the axion isocurvature perturbation is suppressed.

\section*{Acknowledgments}

This work is supported by IBS under the project code, IBS-R018-D1 [KJB] and the Grant-in-Aid for Scientific Research on Scientific Research A (No.26247038 [KH], No.26247042 [KN], No.16H02189 [KH]),
Young Scientists B (No.26800121 [KN], No.26800123 [KH]) and Innovative Areas (No.26104001 [KH], No.26104009 [KH and KN], No.15H05888 [KN]),
and by World Premier International
Research Center Initiative (WPI	 Initiative), MEXT, Japan.
The work of HB is supported in part by the US Department of Energy, 
Office of High Energy Physics.





\end{document}